%% file: Paper.tex
\documentclass[conference]{IEEEtran}

\usepackage{balance}

\usepackage[english]{babel}
\usepackage{times}
\usepackage{xspace}
\usepackage{ifthen}
\usepackage{amsfonts}
\usepackage{amssymb}
\usepackage{graphicx}
\usepackage{xfrac}
\usepackage{array}
\usepackage{color}
\usepackage{xfrac}
\usepackage{fancyvrb}
\usepackage{multirow}

\usepackage{booktabs}

\usepackage{listings,algorithm,algorithmic}

\definecolor{verylightgray}{gray}{0.98}
\usepackage{listings}
\usepackage{courier}
\usepackage[framemethod=TikZ]{mdframed}

\usepackage{booktabs}
\usepackage{amsmath}
\usepackage{algorithm}

\usepackage{caption}
\usepackage{subcaption}

\usepackage{listings}
\usepackage{color}

\definecolor{dkgreen}{rgb}{0,0.6,0}
\definecolor{gray}{rgb}{0.5,0.5,0.5}
\definecolor{mauve}{rgb}{0.58,0,0.82}

\graphicspath{{figures/}}

\newboolean{showcomments}
\setboolean{showcomments}{false}
\ifthenelse{\boolean{showcomments}}{%
   \newcommand{\bnote}[2]{
    \fbox{\bfseries\sffamily \color{blue}{#1}}
    {\sffamily$\blacktriangleright$\color{red}{#2}$\blacktriangleleft$}}
}{%
   \newcommand{\bnote}[2]{}
}

\newcommand{\ie}{i.e.,\xspace}
\newcommand{\eg}{e.g.,\xspace}
\newcommand{\etal}{\emph{et al.}\xspace}

\newcommand{\secref}[1]{Section~\ref{#1}\xspace}
\let\Secref\secref
\newcommand{\figref}[1]{Figure~\ref{#1}\xspace}
\let\Figref\figref
\newcommand{\tabref}[1]{Table~\ref{#1}\xspace}

\newcommand{\equref}[1]{Equation~\eqref{#1}\xspace}

\newcounter{rq}
\newcounter{q}
\newenvironment{Marking}{
	\vspace{+0.1cm}
	\begin{mdframed}[style=MyWhiteBlackFrame]
}
	{\end{mdframed}}

\newcounter{f}
\newcounter{deflist}
\newcounter{mydef}
\newcommand{\myparagraph}[1]{\noindent\textbf{#1}}

\newcommand{\ourtechnique}{LibCUP\xspace}

\newcommand{\davidtechnique}{LibRec\xspace}
\begin{document}


\title{Automated Inference of Software \\Library Usage Patterns}
\author{
    \IEEEauthorblockN{Mohamed Aymen Saied\IEEEauthorrefmark{1}, Ali Ouni\IEEEauthorrefmark{2}, Houari Sahraoui\IEEEauthorrefmark{1}, Raula Gaikovina Kula\IEEEauthorrefmark{2}, Katsuro Inoue\IEEEauthorrefmark{2}, David Lo\IEEEauthorrefmark{3}} 
   \IEEEauthorblockA{\IEEEauthorrefmark{1} DIRO, Universit\'e de Montr\'eal, Montr\'eal, Canada  \\\{saiedmoh, sahraouh\}@iro.umontreal.ca}
    \IEEEauthorblockA{\IEEEauthorrefmark{2} Graduate School of Information Science and Technology, Osaka University, Japan   \\\{ali, raula-k, inoue\}@ist.osaka-u.ac.jp}
    \IEEEauthorblockA{\IEEEauthorrefmark{3} School of Information Systems, Singapore Management  University, Singapore  \\davidlo@smu.edu.sg}
	}

\maketitle

\begin{abstract}

Modern software systems are increasingly dependent on third-party libraries. It
is widely recognized that using mature and well-tested third-party libraries can
improve developers' productivity, reduce time-to-market, and produce more reliable software. Today's open-source 
repositories provide a wide range of libraries that can be freely downloaded and used. However, as software libraries are 
documented separately but intended to be used together, developers are unlikely to fully take advantage of these reuse opportunities. 
In this paper, we present a novel approach to automatically identify third-party library usage patterns, i.e., collections of libraries that are commonly used together by developers. Our approach employs hierarchical clustering technique to group together software libraries based on external client usage. To evaluate our approach, we mined a large set of over 6,000 popular libraries from Maven Central Repository and investigated their usage by over 38,000 client systems from the Github repository. Our experiments show that our technique is able to detect the majority (77\%) of highly consistent and cohesive library usage patterns across a considerable number of client systems.

\end{abstract}



\input{sections/Introduction}

\input{sections/Examples}

\input{sections/Approach}
\input{sections/EmpiricalStudy}


\input{sections/Discussion}

\input{sections/RelatedWork}

\balance

\section{Conclusion}\label{sec:conclusion}

\input{sections/Conclusion}

\bibliographystyle{IEEEtran}
\bibliography{IEEEabrv,bib/local,bib/api}


\end{document}

%% file: sections/Introduction.tex
\section{Introduction}\label{sec:introduction}
Third-party software libraries have become an integral part of modern software development. Today's software systems increasingly depend on external libraries, to reduce development time, and deliver reliable and quality software. Developers can take the benefit of freely reusing functionality provided by well-tested and mature third-party libraries and frameworks through their Application Programming Interfaces (APIs) \cite{Robil09a}. Developers have to cope with the complexity of writing their code from scratch and re-inviting the wheel by automatically identifying existing library usage patterns. 

Much research efforts have been dedicated to the identification of library API usage patterns \cite{Wang:2013:MSH:2487085.2487146,Uddi12,Zhon09,SaiedSaner2015,SaiedICPC2015}. The vast majority of existing works focus on the method level within a single library. Indeed, these approaches assume that the set of relevant libraries is already known to the developer, and it is only the methods in these libraries that are unknown. However, this assumption makes the task to find relevant libraries and understand their usage trend a hard and time-consuming activity. 

Today's code repositories on the Internet provide an increasingly large number of reusable software libraries with a variety of functionalities. Automatically analyzing how software projects utilize these libraries, and understanding the extent and nature of software library reuse in practice is a challenging task for developers.
Indeed, software developers can spend a considerable amount of time and effort to manually identify
libraries that are useful and relevant for the implementation of their software.
Worse yet, developers may even be unaware of the existence of these libraries. 
Developers tend to implement most of their features from scratch instead of reusing functionalities provided by third-party libraries as pointed out by several researchers \cite{Uddi12,Zhon09,Ekok11}.
Therefore, we believe that identifying patterns of libraries commonly
used together, can help developers to discover and choose libraries that may be
relevant for their projects' implementation.


In this paper, we propose a novel approach for mining
Library Co-Usage  Patterns, namely \ourtechnique.
We define a usage pattern of libraries as a collection, with different usage
cohesion levels, of libraries that are most frequently jointly  used  in
client systems.
Our approach adopts a variant of DBSCAN, a widely used density-based  clustering algorithm, to detect candidate library usage patterns based on the analysis of their frequency and consistency of usage within a variety of client systems. Different client systems may use utility libraries (e.g., \texttt{JUnit}, \texttt{log4j}, etc.) as well as domain-specific  libraries (e.g., \texttt{httpclient}, \texttt{groovy}, \texttt{spring-context}, etc.).
Thus, the rationale  behind the distribution on different usage cohesion levels of libraries in a pattern, is to distinguish between the most specific libraries and the less specific ones. Moreover, our approach is intended to be used first to identify patterns of particular libraries that interest a developer.
These libraries could then be fed to existing approaches \cite{Uddi12,Zhon09,SaiedSaner2015,SaiedICPC2015} to recommend particular methods to be used in different contexts. Moreover, \ourtechnique provides a user-friendly visualization tool to assist developers in exploring the different library usage patterns.

We evaluate our approach on a large dataset of over 6,000 popular libraries,
collected from Maven Central repository\footnote{http://mvnrepository.com} and
investigated their usage from a wide range of over 38,000 client systems from
Github repository\footnote{www.github.com}, from different application domains. 
Furthermore, we evaluated the scalability of \ourtechnique as compared to LibRec \cite{david13}, a state-of-the-art library recommendation technique based on  association rule mining and collaborative filtering. We also performed a ten-fold
cross validation  to evaluate the generalizability of the identified usage
patterns to potential new client systems. Our results show that across a considerable variability of client systems, the identified usage patterns by \ourtechnique remain more cohesive than those identified by LibRec. The main contributions of the paper can be summarized as follows:

\begin{enumerate}
\item We introduce a novel approach for mining multi-level usage patterns of libraries using an adapted hierarchical clustering technique. Our  approach is supported by a user-friendly tool 
to visualize and navigate through the identified library usage
patterns \cite{ASEdouble-blind-tool}.


\item We mine a large dataset of over 6,000 popular libraries from Maven
repository and investigated their usage from a wide range of over 38,000 client systems from Github.

\item We evaluate the effectiveness of our approach in terms of cohesiveness and generalizability of the identified patterns. Results show that our approach was able to identify a larger number of usage patterns, on different usage cohesion levels.

\end{enumerate}

The remainder of the paper is organized as follows.
\secref{sec:motivationExamples} motivates the usefulness of \ourtechnique with two
real-world examples. We detail our approach in \secref{sec:approach}. We present our experimental study to evaluate the proposed approach in \secref{sec:Validation}, while providing discussions in  \secref{sec:discussion}. \secref{sec:relatedWork} presents the related work. Finally, we conclude and suggest future work in \secref{sec:conclusion}.

%% file: sections/Examples.tex
\section{Motivation and Challenges}\label{sec:motivationExamples}
In this section, we present two real-world scenarios to motivate the usefulness of library co-usage patterns.
In the first example, the goal is to find a set of libraries that allow to meet the requirements of a given software system.
In the second example, the goal is to decide between two libraries with similar functionalities to be used in a software system.
In this context, we assume that a library with more potential to be used with other related libraries is preferred.
The related libraries are assumed to extend the features of the software system.

\subsection{Learning-Environment Example}\label{sec:sakaiproject}
Let us consider a software development team responsible of the task of maintaining a Web portal for a growing private university with around 4,000 undergraduate and graduate students.
The university is planing to move from a simple Web portal to an advanced course management system to provide adequate service to their students and faculty members.
As a first step, the development team decided to go through an exploratory phase, during which they developed a situational application to assess the turnout rate in the new learning environment.
This application allows students and faculty to schedule activities related to courses and maintain deadlines related to projects.
It should also allow real-time conversations between course or project participants.

Based on these requirements, developers found that their application requires some basic functionalities including a \textit{scheduling} and an \textit{emailing} service that have to be integrated.
In this situation, developers can either implement the different features from scratch, or reuse features provided by existing libraries.
In both cases, they may spend a considerable time and effort for either implementing the features or finding compatible and useful libraries to be integrated in the application.

The development team later find out that they are required to use the {\tt quartz}
 library to implement the scheduler.
With this new constraint, the developers have to solve the following challenges:
\begin{itemize}
\item \textit{What is the recommended emailing library that best complements the {\tt quartz} library?} The selection should take into account assumed compatibility with the  {\tt quartz} library as well as the effort needed to integrate the library into the system.
\item \textit{More generally, what related libraries can be used to implement the remaining features of their software system?} The developers might be interested in related libraries that are commonly used by similar systems with the {\tt quartz} library.
\end{itemize}

Addressing these two challenges could be a complex task for developers if done
manually. Indeed, developers should check in open-source code repositories to find similar projects, and investigate their library usage. Manually finding libraries that are commonly used together in a particular scenario and understanding the current usage practice for a particular library is unlikely to be effective. 

\subsection{ Web Application Frontend Example}\label{sec:gwt}
We now consider another scenario with Aaron, a freelance programmer, who seeks to implement an inventory management web application.
Aaron decided to develop his web applications in industrial setting, where the back-end is implemented in Java and the front-end is implemented in a Java/XML based framework. For the user interfaces, several libraries can be used; the most popular ones are {\tt primefaces}
 the UI component library for Java Server Faces, and {\tt gwt-user}
  of the Google Web Toolkit.

Aaron has to decide which library to use: {\tt primefaces} or  {\tt gwt-user}. In other words:
\begin{itemize}
\item \textit{Which library is the best option in terms of future extension of the software system's functionalities?} Aaron prefers libraries that are usually used with many other libraries, which offers a large variety of functionalities. This provides a high potential of extensions of his software system.
\end{itemize}

In both examples, we consider that mining patterns of libraries used jointly by many client systems may provide insights to make the best decisions.

\subsection{Challenges: Mining Library Usage}\label{sec:cousage}
In this work, we mine the 'wisdom of the crowd' to discover usage patterns of software libraries.
Studying the current library usage within similar systems may provide hints on compatibility and relevance between existing libraries.
 We assume that libraries that are commonly used together are unlikely to have compatibility and integration issues.

The goal is to discover which sets of libraries are commonly used together by similar systems.
To this end, our approach is designed to find multiple layers, i.e., levels,
of relevant libraries according to their usage frequency.
For effective reuse, developers can go through the different levels inside the usage patterns to discover relationships, with different strengths, between the collection of related libraries.

For the first motivating example, we use the usage patterns to discover that {\tt commons-email} library
, which is a popular emailing library, complements the {\tt quartz} library.
Furthermore, by using the multi-layers structure of our patterns, developers can then find related libraries that would complement, at different degrees, both {\tt quartz} and {\tt commons-email}.

For the second motivating example, we found that {\tt gwt-user} library is part of a usage pattern with many other related libraries including {\tt gwt-dev}
, {\tt gwt-servlet}
, {\tt gwt-incubator}
, and {\tt gin}
.
This collection of libraries covers different functionalities such as browser support, widgets, optimization, data binding, and remote communication. All these features are opportunities for future extensions, and we are confident that they can be integrated together as demonstrated by the client systems that already used them.
Conversely, Aaron found that, although {\tt primefaces} library might be useful for his system, it is not widely used with other libraries and, then, does not offer a sufficient guarantee of future integration with other libraries.



These two examples show that the task of identifying library usage patterns becomes more and more complex, especially with the exponentially growing number of libraries available in the Internet.
This motivates our proposal of automatically identify library usage patterns to assist developers in reusing and integrating libraries and, then, increase their productivity.



%% file: sections/Approach.tex
\section{The proposed approach}\label{sec:approach}

In this section, we present our approach, \ourtechnique, for mining library usage
patterns. Before detailing the used algorithm, we provide a brief overview of our approach and describe our visualization technique to explore the identified of library usage patterns.

\subsection{Approach Overview}\label{sec:approachOverview}

Our approach takes as input a set of popular libraries, and a wide variety of
their client systems extracted  from existing open-source repositories. The
output is a set of library usage patterns, each pattern is a collection of
libraries , organized within different layers according to their co-usage frequency.

We define a \textit{library co-usage pattern} (LCUP) as a collection of libraries that are commonly used together. 
A LCUP represents an exclusive subset of libraries, distributed on different usage cohesion layers. 
A usage cohesion layer reflects the co-usage frequency between a set of libraries. 
%
%
%
%

Indeed, similar client systems may share some domain specific libraries, but they may at the same time share some utility libraries which are more commonly used by a large number of systems. For this reason, we seek a technique that can capture co-usage relationships between libraries at different levels. 

Our approach proceeds as follows. First, the input dataset is analyzed to identify the different client systems depending on each library. 
Then, the dependency information is encoded using usage vectors. Indeed, each  library in the dataset is characterized with a usage vector which encodes information about (1) their client systems and (2) the rest of other systems in the dataset that are not using it.
Finally, we use hierarchical clustering technique based on DBSCAN to group the libraries that are most frequently co-used together by clients. 
All libraries that have no consistent usage through the client systems are isolated and considered as noisy data.

\subsection{Clustering Algorithm}

Our clustering is based on the algorithm DBSCAN \cite{Ester96adensity-based}. DBSCAN is a density based algorithm, i.e., the clusters are formed by recognizing dense regions of points in the search space. 
The main idea behind it, is that each point to be clustered must have at least a minimum number of points in its neighbourhood. 
This property of DBSCAN allows the clustering algorithm to filter out all points that are not located in a dense region of points in the search space. 
In other words, the algorithm clusters only relevant points and leaves out noisy points.

This specific property explains our choice of the clustering algorithm DBSCAN to detect usage patterns of libraries. Indeed, not all libraries of the dependency dataset are to be clustered because some are simply not co-used with specific subsets of the libraries, while others are co-used with almost all the subsets of libraries.

\begin{figure}[!t]
\centering
\includegraphics[width=0.95\columnwidth]{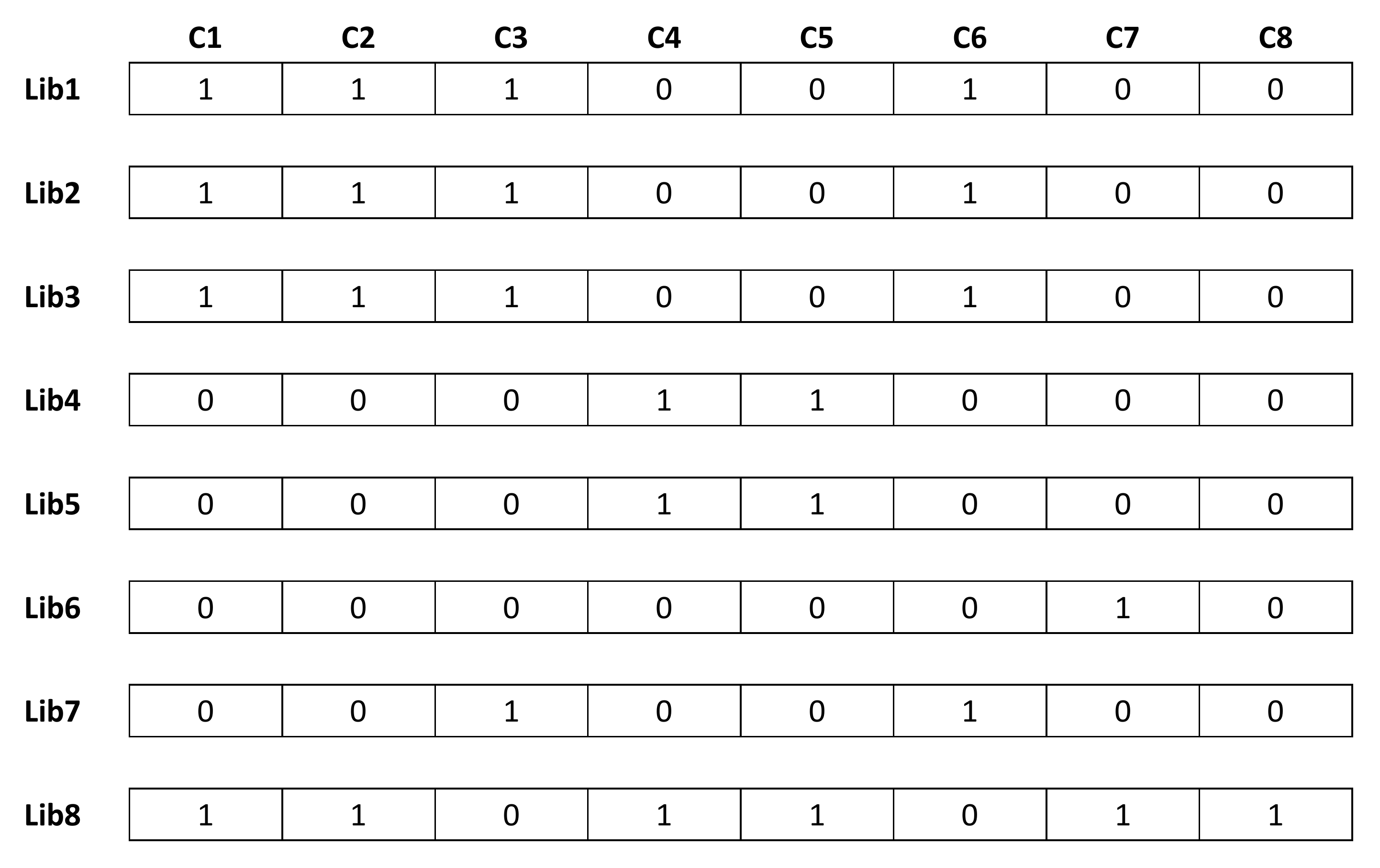}
\caption{The usage vector representing the dependency between eight client programs and eight libraries.}
\vspace{-0.3cm}
\label{fig:methodrep}
\end{figure}

In our approach, each library is represented as a usage vector that has constant length $l$. The vector length is the number of all client programs which use the libraries in the dataset.
\Figref{fig:methodrep} shows that the considered dataset represents 8 client systems depending on 8 third-party libraries. For an external library, $Lib_{x}$, an entry of $1$ (or $0$) in the $i^{th}$ position of its usage vector, denotes that the client system corresponding to this position depends (or does not depend) on the considered library.
Hence, summing the entries in the library's vector represents the number of its client program in the dataset. 
For instance, in \Figref{fig:methodrep}, the usage vector of {\it Lib1} shows that the four client systems {\it C1}, {\it C2}, {\it C3} and {\it C6} depend on this library. We can also see that these systems depend on other libraries including {\it Lib2, Lib3} but  none of them depends on {\it Lib5}.

DBSCAN constructs clusters of libraries by grouping libraries that are close to each other, thus forming a dense region (\ie similar libraries) in terms of their co-usage frequency. 
For this purpose, we define the Usage Similarity, $USim$ in \equref{eq:USim}, between two libraries $Lib_{i}$ and $Lib_{j}$, 
using the Jaccard similarity coefficient with regards to the client programs,
$Cl\_sys$, of $Lib_{i}$ and $Lib_{j}$.
 
The rationale behind this is that two libraries are close to each other (short distance) if they share a large subset of common client systems.    

\vspace{-0.3cm}
\begin{small}
\begin{equation}
USim(Lib_i,Lib_j) =
\frac{ |Cl\_sys(Lib_i) \cap Cl\_sys(Lib_j)|}
		{|Cl\_sys(Lib_i) \cup Cl\_sys(Lib_j)|}
\label{eq:USim}
\end{equation}
\end{small}
\vspace{-0.25cm}

\noindent where $Cl\_sys(Lib)$ is the set of client programs depending on the library $Lib$.
For example, the $USim$ between the libraries {\it Lib1} and {\it Lib6} in \Figref{fig:methodrep} is $\frac{2}{4}$ since these libraries have in total $4$ client programs, and $2$ of them are common for {\it Lib1} and {\it Lib6}. 
The distance between the points in the search space corresponding to two libraries $Lib_{i}$ and $Lib_{j}$ is then computed as $Dist = 1 - USim(Lib_i,Lib_j)$.

DBSCAN requires two parameters to perform the clustering. 
The first parameter is the minimum number of points in a cluster, $minP$. 
We set this parameter at $2$, so that a usage pattern must include at least two libraries of the studied dataset. 
The second parameter is, $epsilon$, the maximum distance within which two points can be considered as neighbor, each to other. 
In other words, $epsilon$ value controls the minimal density that a clustered region can have. 
The shorter is the distance between libraries within a cluster the more dense is
the cluster.

However, it is insufficient merely to apply a generic algorithm like DBSCAN `out of the box'; we need to define problem-specific clustering to provide usage patterns distributed into different levels of usage cohesion. 
 
In the next subsection, we describe our adaptation to the standard DBSCAN to support hierarchical clustering based on different values for $epsilon$ to identify different usage cohesion level in the inferred patterns.

%
%

\subsection{Multi-layer Clustering}

We have introduced a variant of DBSCAN (which we call $\epsilon$-DBSCAN) specifically for the library usage patterns identification problem. 
In DBSCAN, the value of the $epsilon$ parameter influences greatly the resulting clusters. A value of $0$ for $epsilon$, means that each cluster must contain only libraries that are completely similar (\ie distance among libraries belonging to the same cluster must be 0). Our idea is to `relax' the $epsilon$ parameter that controls the  constraints on the requested density within clusters. 

On the one hand, if we set $epsilon$ at fixed small value, \eg $epsilon = 0$, this will produce patterns that are very dense. Consequently, the resulted usage patterns will include only libraries that exhibit a high co-usage score. 
On the other hand, an increase of the $epsilon$ value, \eg $epsilon = 0.3$, will
result in an additional external layer of patterns that exhibit less co-usage
score. Therefore, we iteratively apply the standard DBSCAN at different levels of $epsilon$ in order to have library usage patterns organised as multi-layers, each represents a particular co-usage score.


\begin{algorithm}[t]
\caption{$\epsilon$-DBSCAN: Hierarchical DBSCAN algorithm} 
\small
\label{algo:dbscan}
\begin{algorithmic}[1]
\STATE $\epsilon$-DBSCAN(DataSet, maxEpsilon, MinNbPts, epsilonStep)$\{$
\STATE epsinon $<$-- 0
\WHILE{epsilon $<$ maxEpsilon}
\STATE DBSCAN(DataSet, maxEpsilon, MinNbPts, epsilonStep)
\STATE clusters $<$-- DBSCAN.clusters
\STATE noisyPoints $<$-- DBSCAN.noisyPoints
\STATE compositePoints $<$-- constructPoints(clusters)
\STATE Dataset $<$-- noisyPoints + compositePoints
\STATE epsilon $<$-- epsilon + epsilonStep
\ENDWHILE 
\STATE $\}$
\STATE constructPoints(clusters)$\{$
\FOR{each C in clusters}
\STATE compositePoints $<$-- OR(all points of C)
\ENDFOR
\STATE $\}$

\end{algorithmic}
\vspace{-0.1cm}
\end{algorithm}

As a result, our $\epsilon$-DBSCAN, build the clusters incrementally by relaxing the epsilon parameter, step by step. 
Algorithm \ref{algo:dbscan} shows the pseudo-code of our incremental clustering technique, $\epsilon$-DBSCAN. 
First, $\epsilon$-DBSCAN takes as input a dataset containing all the libraries and their client systems within a specific format, then it cluster them using the standard DBSCAN algorithm with epsilon value of~$0$. This step results in clusters of libraries that are always used together, as well as multiple noisy points left out. For each produced cluster, we aggregate the usage vectors of its libraries using the logical disjunction in one usage vector. Then, a new dataset is formed which includes the aggregated usage vectors and the usage vectors of noisy libraries from the previous run. This dataset is then fed back to the DBSCAN algorithm for clustering, but with a slightly higher value of epsilon, \ie $epsilon = epsilon + epsilonStep$. 
This procedure is repeated in each step until reaching $maxEpsilon$ a maximum
value for epsilon, given as a parameter.

%
%
%

\begin{figure*}
    \centering
    \begin{subfigure}[b]{0.23\textwidth}
        \centering
        \includegraphics[height=0.1\textheight]{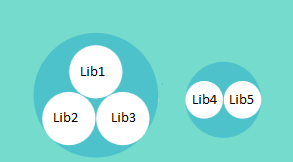}
         \caption{ $epsilon = 0$}
        \label{fig:clusters-a}
    \end{subfigure}
    ~~~~
    \begin{subfigure}[b]{0.2\textwidth}
        \centering
        \includegraphics[height=0.1\textheight]{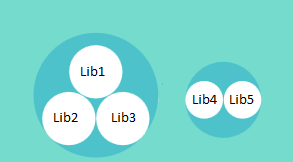}
         \caption{ $epsilon = 0.25$}
        \label{fig:clusters-b}
    \end{subfigure}
    ~~~~~
    \begin{subfigure}[b]{0.23\textwidth}
        \centering
        \includegraphics[height=0.1\textheight]{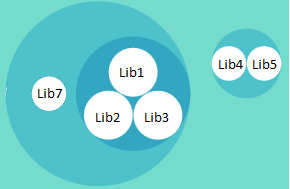}
         \caption{ $epsilon = 0.5$}
        \label{fig:clusters-c}
    \end{subfigure}
    ~
    \begin{subfigure}[b]{0.23\textwidth}
        \centering
        \includegraphics[height=0.1\textheight]{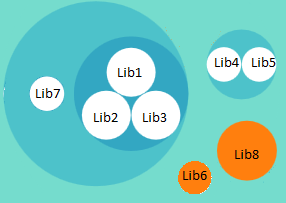}
         \caption{ $epsilon = 0.75$}
        \label{fig:clusters-d}
    \end{subfigure}
    \caption{Resulting clusters of applying the incremental algorithm  $\epsilon$-DBSCAN to the library dataset presented in Figure \ref{fig:methodrep}.}
\label{fig:clusters}
\vspace{-0.5cm}
\end{figure*}

For example, \Figref{fig:clusters} shows the incremental clustering of the libraries in \Figref{fig:methodrep} using $\epsilon$-DBSCAN.  
In this example, the initial dataset contains 8 libraries, {\it Lib1, ..., Lib8}, in which the $epsilon$ parameter is incremented in each step by $epselonStep = 0.25$ with the epsilon maximum value set to $maxEpsilon = 0.55$. 
As shown in \Figref{fig:clusters-a}, the first step produces two clusters at
$epsilon = 0$. The two clusters include respectively ({\it Lib1, Lib2, Lib3})
and {\it Lib4, Lib5}. These libraries are clustered at the most dense level since, in each cluster,  these libraries were frequently co-used together.
The second step is performed with $epsilon = 0.25$ as illustrated in \Figref{fig:clusters-b}. 
For this step, there is no change in the dataset since the distances are larger than 
the current $epsilon$ value. Finally at $epsilon = 0.5$, as illustrated in
\Figref{fig:clusters-c} a new cluster involving 2 density level is generated.
This cluster includes {\it Lib7} in addition to ({\it Lib1, Lib2, Lib3}) since they share 2 out of the 4 common client systems. 

We can notice that {\it Lib6} is a rarely used library and {\it Lib8} 
is a utility library used with almost all the considered client systems, showing
no particular usage trend. Thus at the last iteration of $\epsilon$-DBSCAN 
illustrated in \Figref{fig:clusters-d}, the libraries {\it Lib6} and {\it Lib8}
are left out as  noisy points since their distance from the clustered libraries is larger than the maximum epsilon value, which is $0.55$.

%% file: sections/EmpiricalStudy.tex
\section{Empirical study}\label{sec:Validation}

In this section, we present the results of our evaluation of the proposed approach, \ourtechnique. 
Our study aims at assessing whether \ourtechnique can 
detect \emph{usage patterns of libraries} that are (\textit{i})  \textit{cohesive} enough to provide valuable information to discover relevant libraries, and (\textit{ii}) \textit{generalizable} for new client systems. We also compare the results of our technique \ourtechnique to the available state-of-the-art approach, \davidtechnique \cite{david13}. \davidtechnique combines association rule
mining and collaborative filtering to recommend libraries based on their client usage. For each experiment in this section, we present the
research questions to answer, the research method to address them, followed by the obtained results.

All the material used to run our three experiments is publicly
available in a comprehensive replication package \cite{ASEdouble-blind-tool}.

\subsection{Data collection}\label{sec:dataset}

To evaluate the feasibility of our approach on real-world scenarios, we carried
out our empirical study on a large dataset of Open Source Software (OSS)
projects. As we described earlier, our study is based on widely used libraries
collected from the popular library repository Maven and a large set of client
systems collected from Github repository. Since Github is the host of varying
projects, to ensure validity of quality Github projects, we performed the following filtering on the dataset:

\begin{itemize}
\item \textit{Commit size}. We only included java projects that had more than 1,000 commits.

\item \textit{Forks}. We only include projects that are unique and not forks of other projects.

\item \textit{Maven dependent project}. .   We only included projects that  employ the maven build process (use \texttt{pom.xml} configuration file).
\end{itemize}

Each Github repository may contain multiple projects, each having potentially
several  systems. Each of these systems is dependent on a set of maven libraries, that are defined in a\texttt{ pom.xml} file within the project.

\begin{table} [h]
\centering
\caption{Dataset used in the experiment}
\label{tab:minedLib}
\begin{tabular}{@{}lcccc@{}}
\toprule
\multicolumn{1}{c}{} & \multicolumn{1}{c}{Dataset} \\\midrule
Snapshot Date & 15th January 2015\\
\# of github systems & 38,000 \\
\# of unique dependent libraries &  6,638 \\
\bottomrule
\end{tabular}
\end{table}

Note that for all data, we first downloaded an offline copy of the original software projects (the source code) from Github and the libraries (the jar files) from Maven before extraction. Thereafter, for each library, we selected the latest release. In the beginning we started with 40,936 dependent libraries. However, to remove noise, we filtered out libraries having less than  50 identifiers based on methods, attributes and classes. This process removed libraries that we assume very small or partial copies  of their original libraries and thus are not relevant. Our dataset resulted in 6,638  Maven libraries extracted from unique 38,000 client systems from Github.

The dataset is a snapshot of the projects procured as of 15th January 2015. Our
dataset is very diversified as it includes a multitude of libraries and software
systems from different application domains and different sizes. Overall, the
average number of used libraries per system is 10.56, the median is 6.


%


\subsection{Sensitivity Analysis}
As a first experiment, we evaluated the sensitivity of the patterns' quality, identified by \ourtechnique, with respect to different settings including the dataset size and $maxEpsilon$ values. We aim at addressing the following research question.

\begin{description}
\item[\textbf{RQ1.}] \textit{What is the impact of various experimental settings on the patterns'  quality?}
\end{description}

\subsubsection{Analysis method}\label{sec:setup}
To address \textbf{(RQ1)}, we need to evaluate whether the detected patterns are cohesive enough to exhibit informative co-usage relationships between specific libraries. 
Hence, we use a cohesion metric namely, Pattern Usage Cohesion metric (PUC), to capture the cohesion of the identified patterns. 

PUC is inspired from Perepletchikov et al. \cite{Pere07b} and was originally used to assess the usage cohesion of service interfaces.
It evaluates the co-usage uniformity of an ensemble of entities, which corresponds, in our context, to a set of libraries in the form of a library usage pattern.
PUC values are the range [0,1]. The larger the value of PUC is, 
the better the usage cohesion, i.e., a usage pattern has an ideal usage cohesion (PUC = 1) if all the library patterns are always used together. Let \textit{p} is a library usage pattern, then its PUC is defined as follows:
\begin{equation}
PUC(p) = \dfrac{\sum_{cp} ratio\_used\_Libs(p,cp)}{\vert
C(p)\vert} \in [0,1]
\label{eq:PUC}
\end{equation}
\noindent where \textit{cp} denotes a client system of the pattern \textit{p}, \textit{ratio\_used\_Libs(p, cp)} is the ratio of libraries that belong to the pattern \textit{p} and that are used by the client system~\textit{cp}, and $C(p)$
 is the set of all client systems of all libraries in \textit{p}.

We answer  our first research question \textbf{(RQ1)}, we perform two studies. 

\begin{itemize}
\item \textbf{Study 1.A.} We apply \ourtechnique to our collected dataset described in \secref{sec:dataset}. Then, we investigate the impact of different $maxEpsilon$ values on the PUC results of the detected patterns. 

\item \textbf{Study 1.B.} We investigate the scalability of our technique. We fix the $maxEpsilon$ value and we run \ourtechnique several times 
while varying the dataset size to observe the patterns cohesion and the time efficiency.
\end{itemize}

\vspace{0.2cm}
\subsubsection{Results for RQ1}\label{sec:setup}  
\begin{figure}[!t]
\centering
\includegraphics[width=0.9\columnwidth]{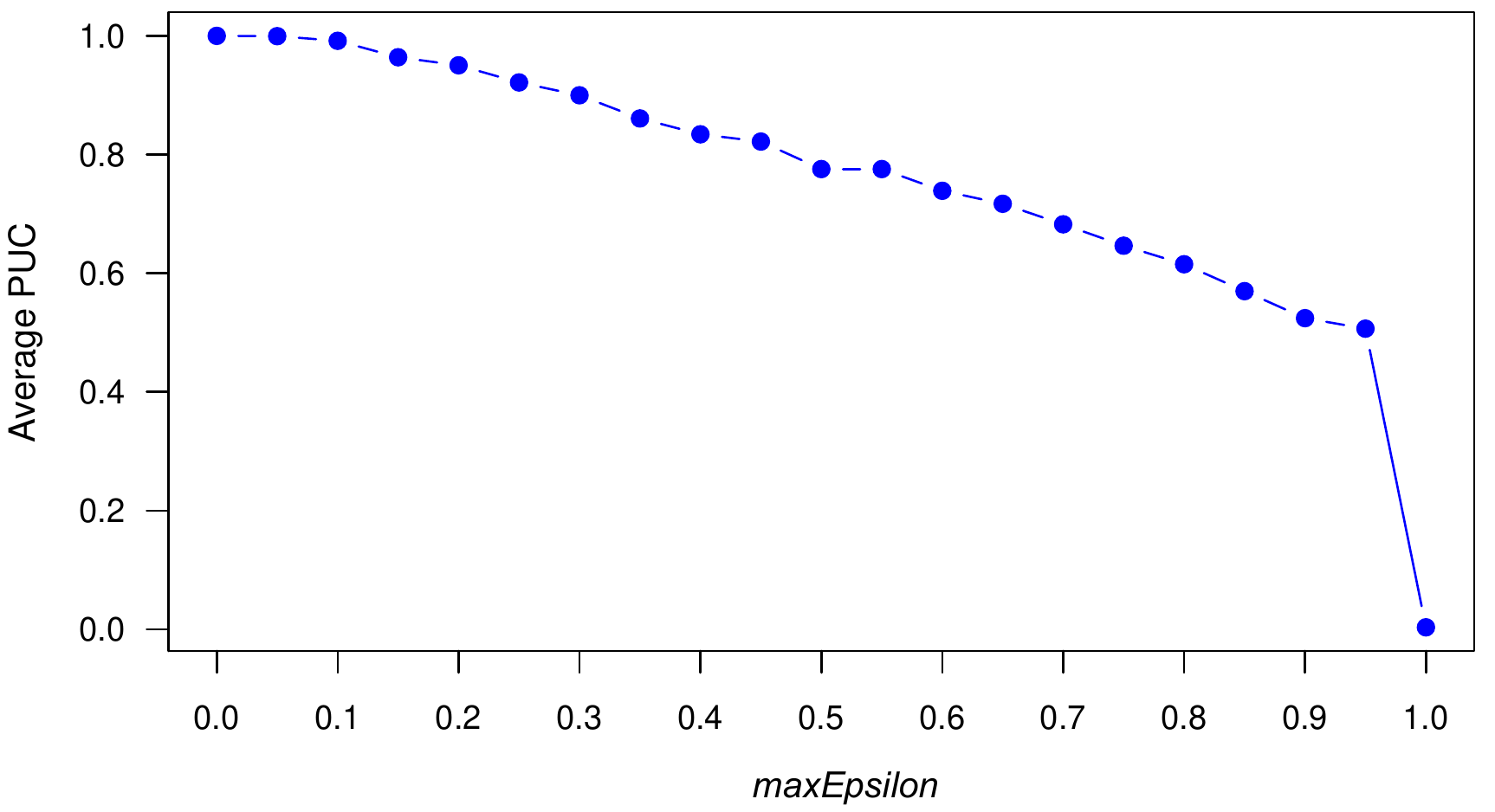}

\caption{Effect of varying $maxEpsilon$ parameter on the average cohesion of the
identified patterns.}

\label{fig:epsilonVScoh}
\vspace{-0.6cm}
\end{figure}

\begin{figure}[!t]
\centering
\includegraphics[width=0.9\columnwidth]{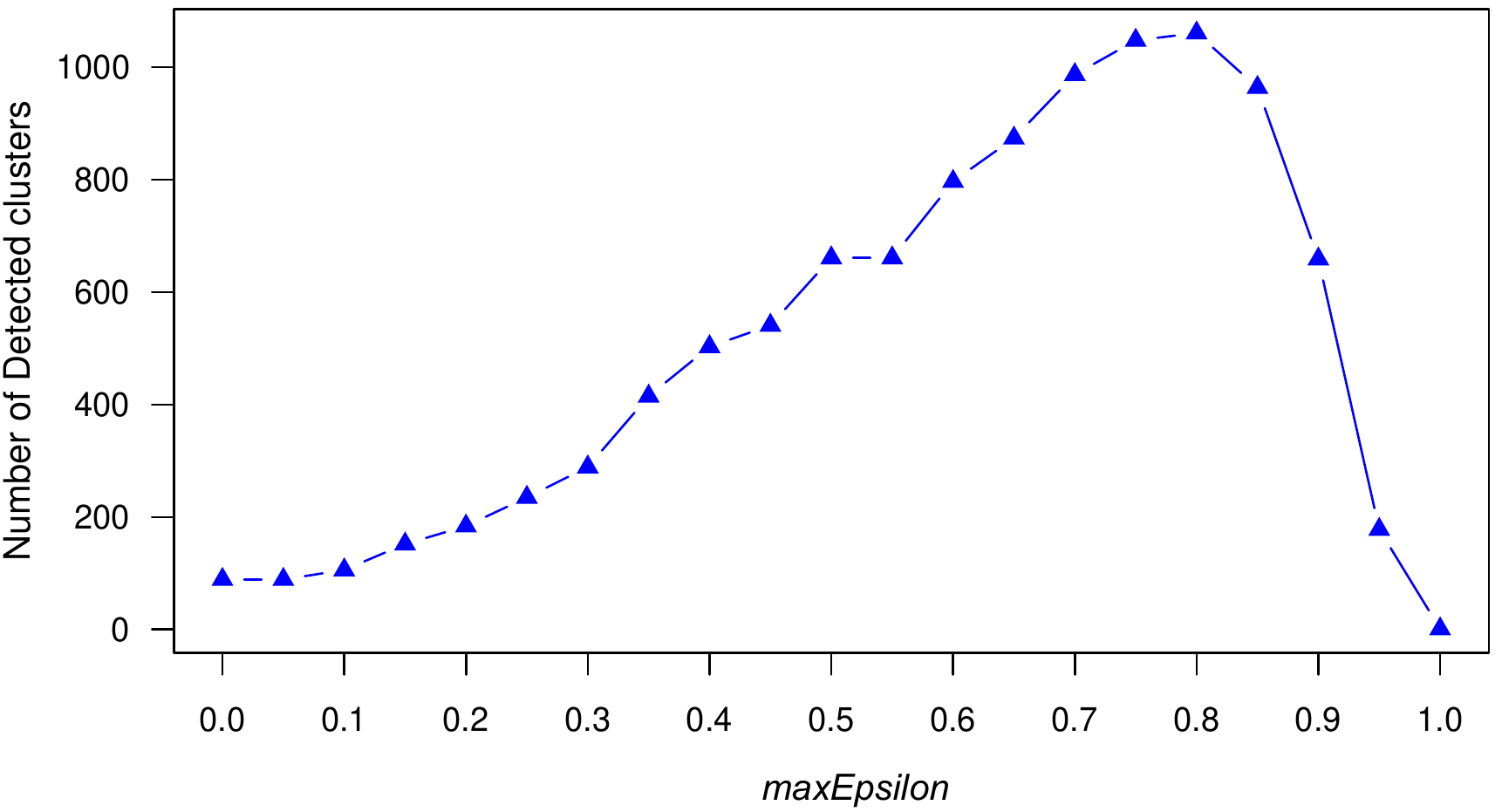}

\caption{Effect of varying \textit{maxEpsilon} parameter on the number of identified patterns.}
\vspace{-0.4cm}
\label{fig:epsilonVSnbPatterns}
\end{figure}

\begin{figure}[!t]
\centering
\includegraphics[width=0.9\columnwidth]{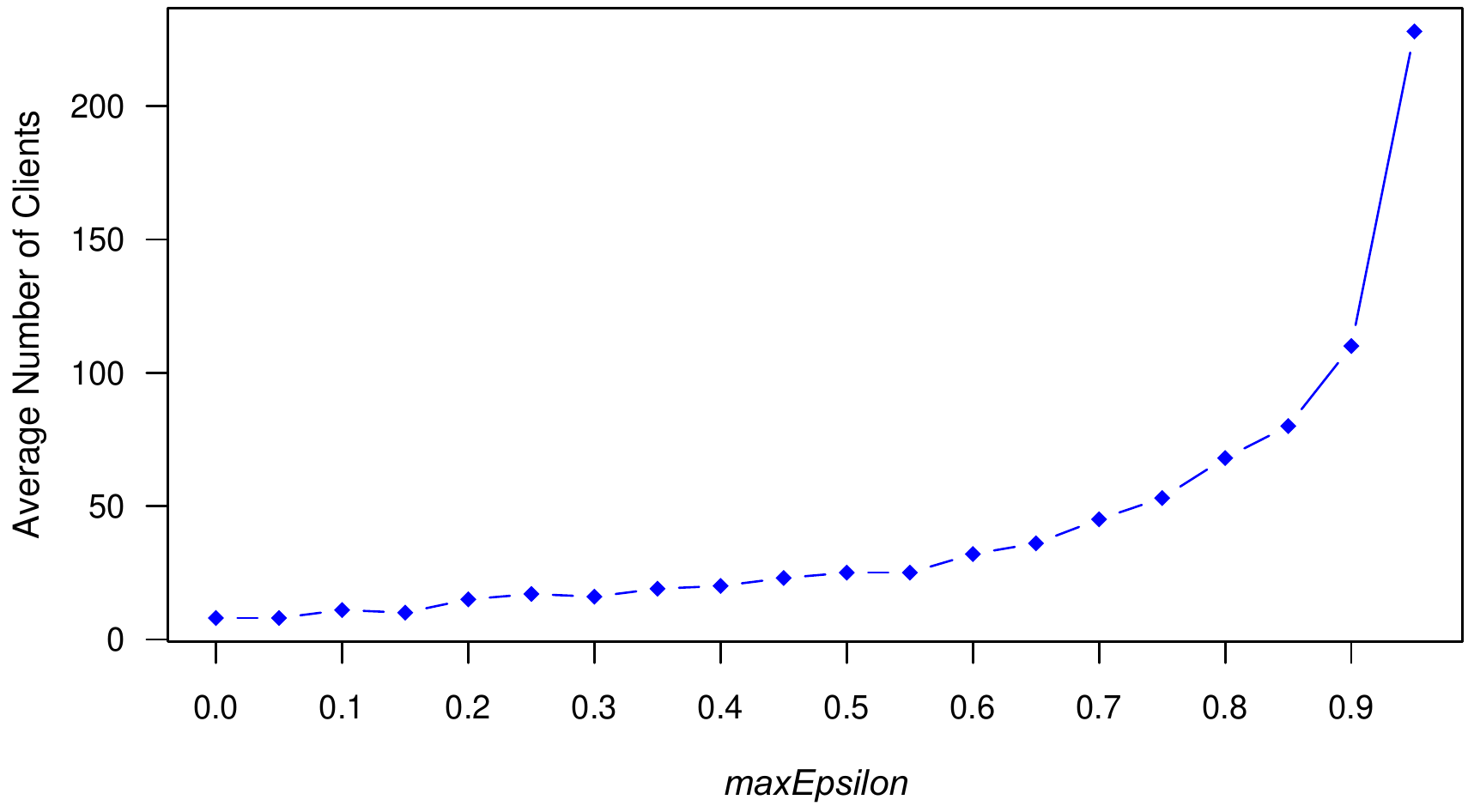}
\caption{Effect of varying $maxEpsilon$ parameter on the average number of clients per pattern.}
\vspace{-0.6cm}
\label{fig:epsiloneVSnbClients}
\end{figure}

The obtained results are as follows.\\
\textbf{\underline{Study 1.A: Sensitivity to \textit{maxEpsilon} parameter.}}
Figures \ref{fig:epsilonVScoh}, \ref{fig:epsilonVSnbPatterns}, and \ref{fig:epsiloneVSnbClients} report the effect of different $maxEpsilon$  values. Our experiments show that the $maxEpsilon$ parameter influences different characteristic of the inferred patterns including the pattern usage cohesion, the number of inferred patterns, and the patterns size.  \Figref{fig:epsilonVScoh} shows that the average PUC ranges from 1 to 0.5, while varying the $maxEpsilon$ in the range [0,0.95]. We notice that even when the $maxEpsilon$ reaches high values, the inferred patterns maintain  acceptable cohesion values. This is due to the incremental construction of the patterns that generates multiple layers of libraries, each reflected at a different density level. Thus, layers inferred at the early steps influence the overall cohesion of the final pattern. 

When $maxEpsilon$ is set to 1 the patterns'  cohesion drop down to 0. This  is because  in the last step all the libraries are clustered into one usage pattern as depicted in \Figref{fig:epsilonVSnbPatterns}.
Moreover, we can clearly see from this figure that the number of inferred patterns increases to reach a peak of 1,061 when $maxEpsilon$ is set to 0.80. In more details, we observe that:

\begin{itemize}
\item Before the peak, some of the existing patterns are enriched with new
libraries to add new external layers to the original usage pattern. Moreover, we
 also noticed that  some new patterns are identified with libraries that were
considered as noisy. This is since we are tolerating less density within clusters when $maxEpsilon$ increases. 
 We observe that this has an effect to increase the global number of inferred patterns. 

\item After the peak, some of the existing patterns are merged without losing their internal structure. 
This result, in turn has an effect to reduce the overall number of inferred patterns.
\end{itemize}

To get more qualitative sense of the obtained results, we noticed that for the
low values of $maxEpsilon$ and up to intermediate values (i.e., 0.5, 0.6),
 the inferred patterns tend to mainly cover domain specific libraries (e.g., program analyzers {\tt jdepend}
 , graphics manipulation {\tt batik}
 , etc.). Those patterns are characterized with an average number of client systems that do not exceed 50 clients per pattern. The more the $maxEpsilon$ parameter is relaxed, the more the patterns are enriched with other  libraries. Starting with specific libraries, the patterns reach a step in which they become enriched  with utility or more generic libraries such as {\tt JUnit}
  and {\tt log4j}
    . For sake of simplicity, we do not present in \Figref{fig:epsiloneVSnbClients} the last step where all the libraries are clustered into one single usage pattern with a larger number of client systems. Based on these results, \ourtechnique uses a $maxEpsilon$ threshold of 0.5 as a defaults parameter.

\textbf{\underline{Study 1.B: Sensitivity to the dataset size.}}
To carry out this experiment, we set the $maxEpsilon$ value to 0.5. This is a
proactive choice to ensure that libraries appearing in the same pattern are used more frequently together than separately. 
Thereafter, we run \ourtechnique with different dataset sizes. In each run, we augmented  the previously used dataset with 1000 libraries, 
and we observed the average cohesion of patterns as well as the execution time taken to infer them. All experiments were carried out on 
a computer with an Intel core i7-4770 CPU 3.40 GHz, with 32 GB RAM.

\begin{figure}[]
\centering
\includegraphics[width=1\columnwidth]{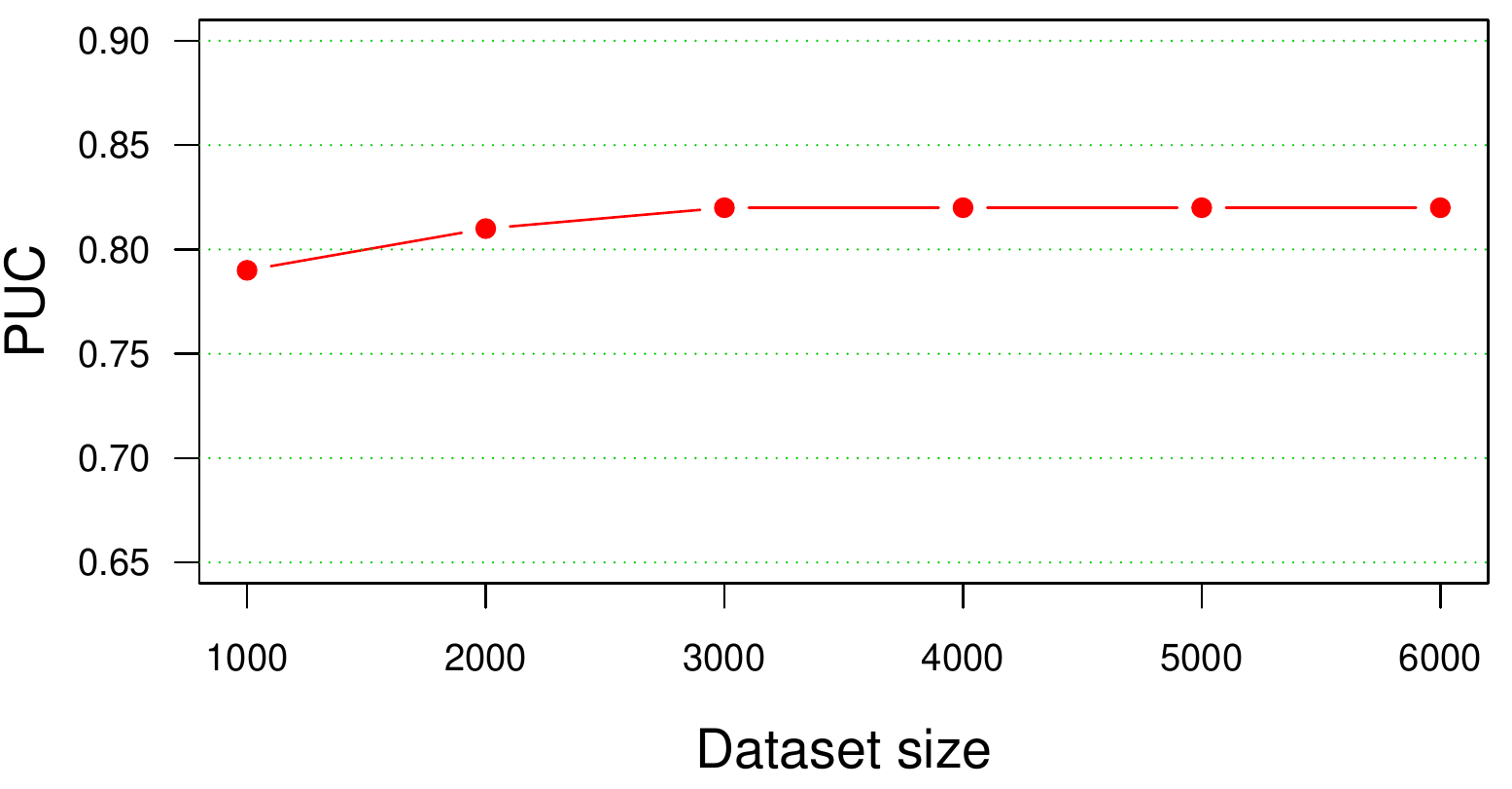}
\caption{Effect of varying the dataset size on the average pattern cohesion $maxEpsilon = 0.5$.}
\label{fig:datasetVSpuc}
\vspace{-0.4cm}
\end{figure}

\Figref{fig:datasetVSpuc} depicts the obtained results for this experiment. We noticed from the figure, that the shape of the graph is consistent for the different dataset size. 
The PUC score slightly increases from 0.79 to stabilize at  0.82 for the last three runs. 
In more details, we found that there is an increase in terms of the number of
inferred patterns from 62 in the first run with an average size of 3 libraries
per pattern, to reach the bar of 500 patterns at the last run with an average
size of 5.5 libraries per pattern. These  results confirm that when considering more libraries, \ourtechnique  is able to enrich the inferred patterns with new libraries while detecting new patterns.

\begin{figure}[]
\centering
\includegraphics[width=1\columnwidth]{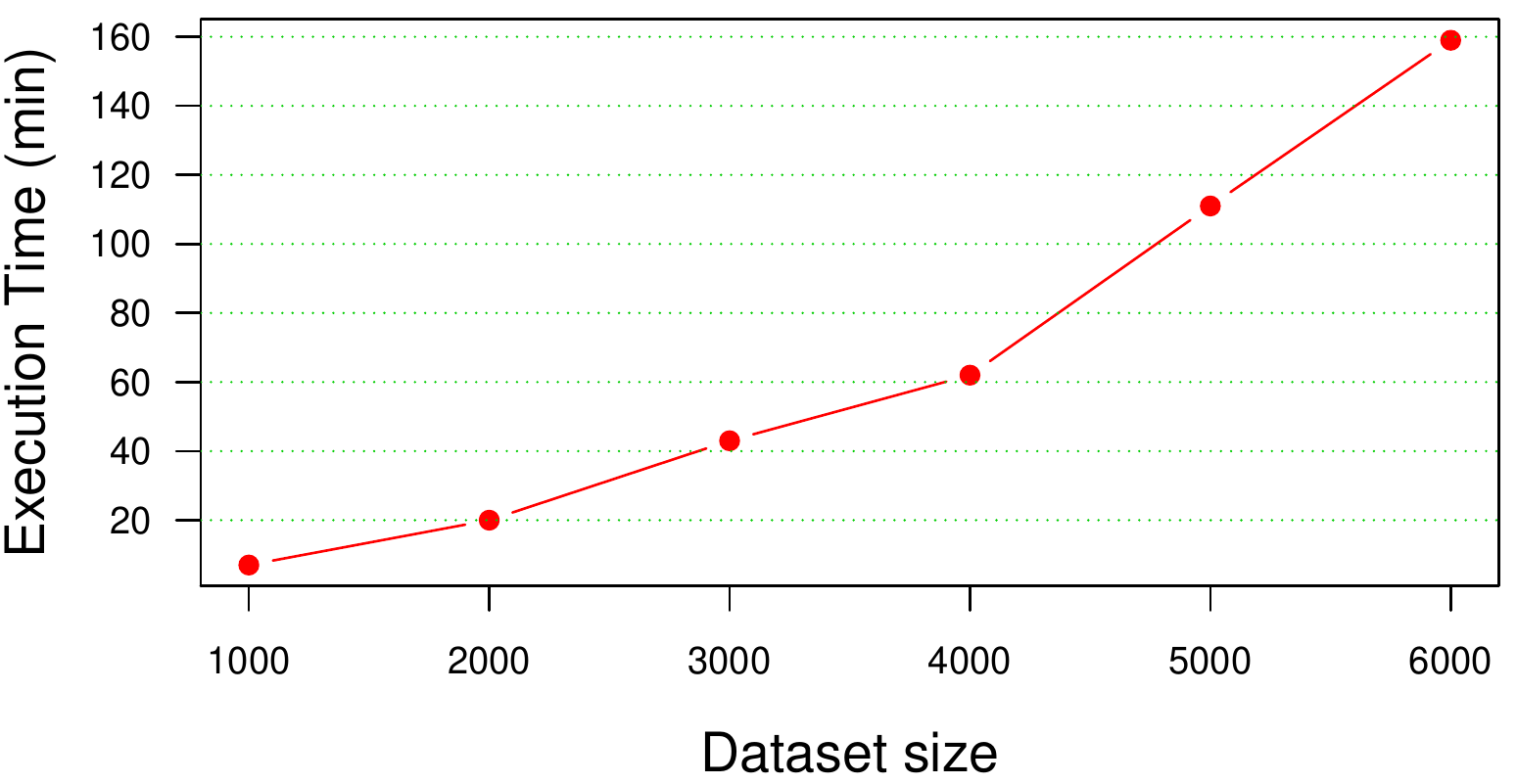}
\caption{Effect of varying the dataset size on the time efficiency with $maxEpsilon = 0.5$.}
\label{fig:datasetVSTime}
\vspace{-0.3cm}
\end{figure}

In terms of time efficiency, \Figref{fig:datasetVSTime} depicts the influence of
the dataset size on the execution time. As it can be seen on the figure, the
execution time of  \ourtechnique  is sensitive to the dataset size, as  expected. At the first run, \ourtechnique took less than 7 minutes to mine a set of 1,000 libraries, while reaching 159 minutes of execution time to mine the large set of 6,000 libraries with their 38,000 client systems. However, it is worth saying  that even with 159 minutes 
of execution time, \ourtechnique can be considered time efficient, since the inference process is done off-line once, then the identified patterns can be easily explored using our interactive visualization tool.

In summary, the obtained PUC results of the identified usage patterns provide evidence
that \ourtechnique exhibits consistent cohesion using our adopted $\epsilon$-DBSCAN technique. Using a default $maxEpsilon = 0.5$, we found that at least 50\%, and up to 100\%, of the usage patterns are co-used together with high PUC. Moreover, our technique is stable and time efficient when varying the size of the mined library set.

\subsection{Evaluation of patterns cohesion}\label{sec:evaluation-cohesion}
As a second experiment, we conduct a comparative study to evaluate the cohesiveness of the identified library usage patterns against the state-of-the-art approach \davidtechnique \cite{david13}. To the best of our knowledge, \davidtechnique \cite{david13} is the only existing approach that has addressed this problem. We aim at addressing the following research
question.

\begin{description}
\item[\textbf{RQ2.}] \textit{To which extent are the identified library usage patterns cohesive as compared to those inferred with \davidtechnique ?}
\end{description}

\subsubsection{Analysis method}\label{sec:setup}
To address our second research questions \textbf{(RQ2)} we conducted a
comparative evaluation of our approach with \davidtechnique in order to better position our approach and
characterize the obtained results.

To infer usage patterns, \davidtechnique is based on  mining association rules obtained from closed itemsets and generators using the Zart algorithm \cite{szathmary2006zart}.  
We applied both \ourtechnique and \davidtechnique to all the selected libraries
of our dataset (cf. \secref{sec:dataset}). Then, we compare the identified usage
patterns of both approaches in terms of PUC. More specifically, we  compare  the
average PUC  values  for all  detected   patterns  of each approach. For
\ourtechnique, we fixed the $maxEpsilon$ value to 0.5 as explained earlier. For \davidtechnique we fixed the $Minconf$ to 0.8 , the $Minsup$ to 0.002 and the $Number$ $of Nearest Neighbors$ to 25 \cite{david13}.

\vspace{0.2cm}
\subsubsection{Results for RQ2}\label{sec:setup}

\tabref{tab:avgCohaymenAndDavid} reports the obtained results for \textbf{RQ2}. On average,
\ourtechnique achieves an average PUC score of 0.82 which outperforms \davidtechnique that was only able to achieve  0.72 of PUC. 
The achieved PUC values  by \ourtechnique reflect high co-usage
relationships between the pattern's libraries making them more cohesive. 

In terms of number of inferred patterns, we observe that our
multi-layer clustering technique allows detecting a reasonable number of
patterns of 531, with a medium size of libraries distributed on the different
layers (i.e., 5.5). On the other hand, \davidtechnique inferred   an abundant number
of patterns up to 3,952, even though it relies on closed itemsets and generators  to construct a compact set of association rules \cite{david13}. Indeed, the set of patterns obtained from the closed itemsets and generators is supposed to be much smaller than the complete set of rules.
However, in practice the inferred patterns with \davidtechnique tend to be many but with smaller size (on average, it generates 2 libraries per pattern). We believe that this large number of small size library patterns will in turn limit the practical adoption and usefulness of the LibRec approach. 

Furthermore, we studied the number of clients per pattern. We noticed from the results of Table \ref{tab:avgCohaymenAndDavid}, that the patterns inferred by \ourtechnique are used on average with 30 client systems. Indeed, by manually investigating these client systems, we found that they generally share common domain specific features. 
For \davidtechnique, the inferred  patterns are used within an excessive number of client systems that share pairs or triplets of libraries which 
are, in most of the cases, utility libraries such as {\tt JUnit},  {\tt log4j},   
{\tt slf4j-api},  {\tt commons-lang} and several others. These libraries are likely to be used by several unrelated client systems.

\begin{table}[h!]
\tabcolsep=0.4cm
\centering
\caption{Average cohesion and overview of the inferred usage patterns for 
\ourtechnique and \davidtechnique.
\label{tab:avgCohaymenAndDavid}}
\begin{tabular}{@{}lcc@{}}
\toprule
\textbf{Number of patterns} & \textbf{\ourtechnique} & \textbf{LibRec} \\ \midrule
Avg PUC                     & 0.82                   & 0.72            \\
Nb Patterns                  & 531                    & 3,952            \\
Avg pattern size            & 5.5                    & 2.0             \\
Nb Clients per Pattern       & 30                     & 2,269            \\ \bottomrule
\end{tabular}
\end{table}


\subsection{Evaluation of patterns generalization}\label{sec:evaluation-cohesion}

In this study, we aim at evaluating whether the identified library usage
patterns with \ourtechnique  can be generalizable in comparison with those of 
\davidtechnique. We aim at addressing the following research question.

\begin{description}
\item[\textbf{RQ3.}] \textit{To which extent are the detected usage patterns generalizable to other  ``new'' client systems, that are not considered in the training dataset?}
\end{description}

\subsubsection{Analysis method}\label{sec:setup}



To answer \textbf{RQ3}, we investigate whether the detected patterns will have
similar PUC values in the context of new client systems. 
 We assume that 
 \textit{detected patterns are said ``generalizable'' if they remain characterized
by a high usage cohesion degree in the contexts of various client systems.}

To evaluate the generalizability of the detected patterns, we perform a ten-fold cross-validation on all the client systems in the dataset. 
We randomly distribute the dataset into ten equal-sized parts. Then, we perform ten independent runs of both approaches, \ourtechnique and \davidtechnique. Each run uses nine parts as training client systems 
to detect possible patterns, and leaves away the remaining part as a validation dataset.
 
The results are sorted in ten runs, where each run has its associated patterns, and its corresponding training and validation client systems. 
Then, we address \textbf{(RQ3)} through two experimental studies as follows.

 \textbf{\underline{Study 3.A.}} We evaluate the cohesion of the detected patterns (as measured by PUC) in the context of validation datasets. In a given run, it is possible that some detected patterns contain only 
libraries that are never used in the validation client systems.
Consequently, to evaluate the generalizability of the detected patterns in each run, we consider only the patterns that contain at least 
one library that is actually used by the run's validation client systems. 

We call such patterns the `\textit{eligible patterns}' for the validation client systems. 
An eligible pattern will have a low PUC if only a small subset of its libraries
is  used by the validation client systems, while the other libraries have not
been used.  As a consequence, it will be considered as ``non-generalizable''.
This study aims at comparing  the PUC results obtained for the training client
systems context and validation client systems context for both \ourtechnique and \davidtechnique.

 \textbf{\underline{Study 3.B.}} 
In this study, we push further the comparison, as \davidtechnique is
specifically designed for library recommendation. We attempt to evaluate whether our approach is also useful in a recommendation context. To this end, we define for the library patterns inferred by \ourtechnique an ad-hoc ranking score based on the pattern cohesion and the library usage similarity. 

For each fold, we identify a recommendation set of useful libraries for
the validation client systems. For each system, we drop half of its libraries and use them as the ground truth. The remaining half is used as input to the recommendation process.  This methodology was also used in \cite{david13} and mimics the scenario where a developer knows some of the useful libraries but needs assistance to find other
relevant libraries. 

For each system that should receive library
recommendation, we first identified potentially useful patterns containing at
least one library from the ground truth set.
Thereafter, we rank the libraries of these patterns according to their recommendation score as defined below:
 \vspace{-0.1cm}
 \begin{equation}
RecScore(L)=\max_{i} \lbrace   USim(L,Lib_i)    /  Lib_i  \in GT  \rbrace
\end{equation}

 where $USim$ is the Usage Similarity in \equref{eq:USim}, and $GT$ is the set
 of libraries conserved as ground truth of the client system that should receive library recommendations.

We  evaluate the ranking for both \ourtechnique and \davidtechnique using two
metrics commonly used in recommendation systems for software
engineering \cite{Avazpour2014,tantithamthavorn2013mining,tantithamthavorn2013using}:
(\textit{i}) the recall rate@$k$, and (\textit{ii}) the Mean reciprocal rank (MRR) as follows. To measure the recall@$k$, we consider $N$ target systems $S$that should receive library recommendations.
For each system $S_i \in S$, if any of the dropped libraries is found in the top-$k$ list of recommended libraries, we count it as a hit. 
The recall rate@$k$ is measured by the ratio of the number of hits
over the total number $N$  of considered systems. Inspired by the previous studies \cite{Avazpour2014,tantithamthavorn2013mining,tantithamthavorn2013using}, we choose the $k$ value to be 1, 3, 5, 7, and 10. Formally, the recall rate@$k$ is defined as follows:
 
\begin{small}
\begin{equation}
Recall~rate@k(S)=\dfrac{\sum\limits_{\forall S_i \in S}isCorrect(S_i,\textrm{Top-}k)}{N}
\end{equation}
\end{small}

\vspace{-0.4cm}
\noindent where the function $isCorrect(S_i,\textrm{Top-}k)$ returns a value of 1 if at least one of the top-$k$ recommended libraries is in the ground truth set, or 0  otherwise.

The MRR  is a statistic measure that is commonly used to evaluate
recommendation systems.  Let $N$ be the number of systems that should receive recommendations, and $rank_i$ is the rank of the first relevant recommendation for the $i^{th}$ system, then the MRR score is calculated as follows:
\begin{equation}
MRR=\dfrac{1}{N}\sum_{i=1}^{N}\frac{1}{rank_{i}}
\end{equation}

\subsubsection{Results for RQ3}\label{sec:setup}
The obtained results are as follows.

\begin{figure}[!t]
\centering
\includegraphics[width=0.85\columnwidth]{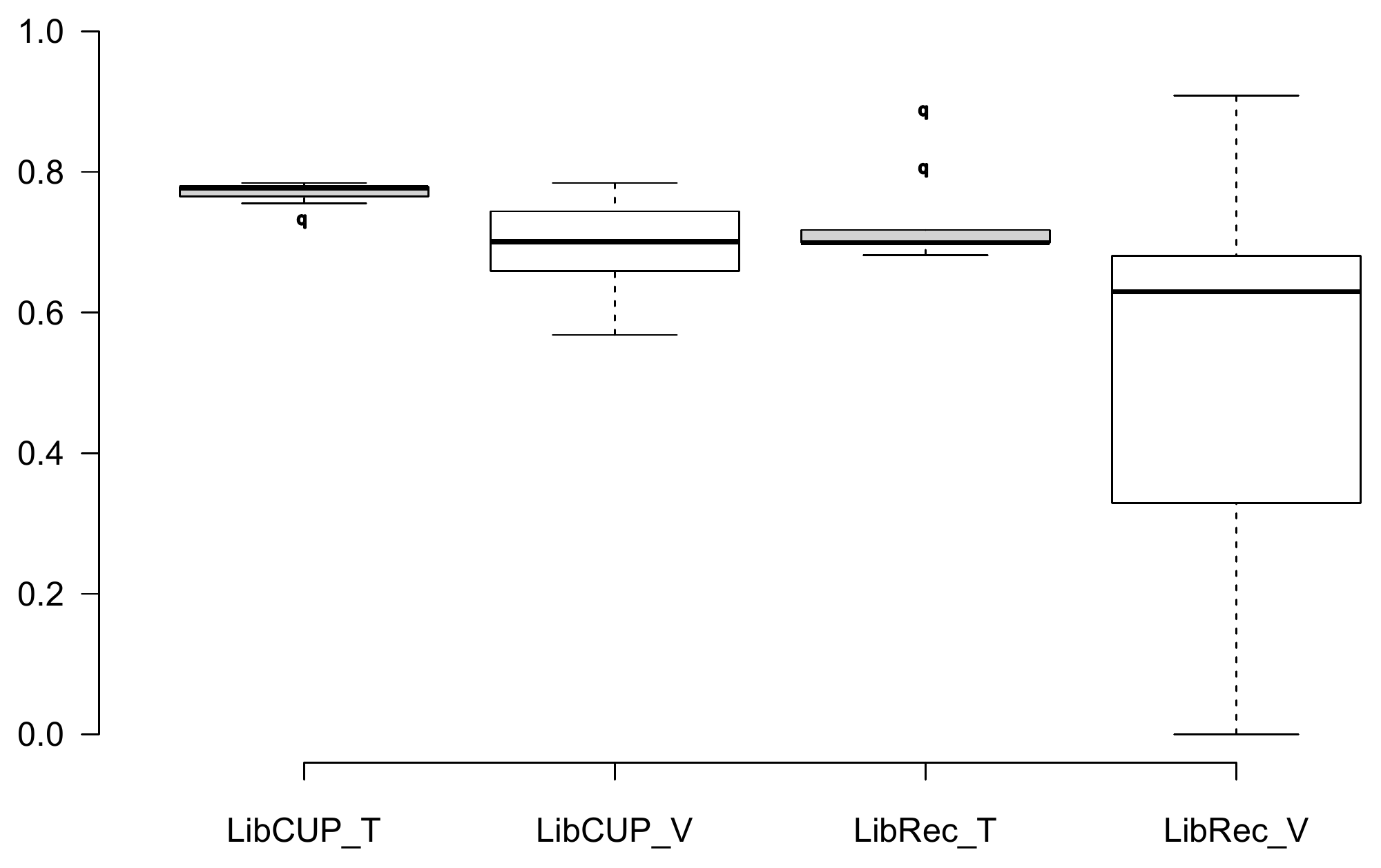}
\caption{PUC results of the identified library usage patterns in the contexts of training (T) and validation (V) clients achieved by each of \ourtechnique and \davidtechnique.}
\label{fig:Cohesion}
\vspace{-0.6cm}
\end{figure}

\textbf{\underline{Study 3.A: Patterns generalizability.}}
To assess the PUC score variation between the training and validation client systems, we first analyze the average value of their corresponding scores collected from all cross-validation runs. Then, we analyze the distribution of the collected values using the median. The results of this study are summarized in \tabref{tab:TrainingvsNCBValidationCoh} and \figref{fig:Cohesion}.

\begin{table}[!t]
\setlength{\belowrulesep}{0.4pt}
\setlength{\aboverulesep}{0pt}
\begin{center}
\caption{Average Training and Validation Cohesion of identified usage 
patterns for \ourtechnique and \davidtechnique. \label{tab:TrainingvsNCBValidationCoh}}

\begin{tabular*}{1\linewidth}{ @{\extracolsep{\fill}} l||c|c||c|c}

\multicolumn{1}{c}{}				&	\multicolumn{2}{c}{{\bf \ourtechnique}} &\multicolumn{2}{c}{{\bf \davidtechnique}}\\

				\cmidrule(l{.05em}r{0.7em}){2-3}	
				\cmidrule(l{.05em}r{0.7em}){4-5}

{\bf }		&	{\bf Training}	&	{\bf Validation}	&	{\bf Training}		&	{\bf Validation}\\
{\bf PUC}	&	{\bf context}	&	{\bf context} &	{\bf context}		&   {\bf context} \\

\toprule
Avg 			&	0.77		&	0.69			&	0.72	&	0.54\\
\midrule
Max 			&	0.78		&	0.78			&	0.88 	&	0.89\\
\midrule
StdDev			&	0.01		&	0.05			&	0.06	&	0.27\\
\bottomrule
\end{tabular*}
\end{center}
\vspace{-0.6cm}
\end{table}

\tabref{tab:TrainingvsNCBValidationCoh} summarizes the PUC results of the detected patterns in the contexts of training and validation client systems. 
In the training context, we notice that the average values is high for both \ourtechnique and \davidtechnique with respectively 77\%  and 72\% of the patterns libraries 
that are co-used together. 
For \ourtechnique, a slight degradation of the PUC value is observed in the
context of validation client systems. We also notice that the standard deviation values 
are very low (0.01 and 0.05). These results reflect that, overall, the 
detected patterns had good PUC in the context of both validation and training client systems. However, for \davidtechnique the
achieved average PUC values are significantly lower in the validation context comparing to the training context.

In more details, the distribution of PUC values for all detected usage patterns
in \figref{fig:Cohesion} confirms the above-mentioned finding. Indeed, the
medians and lower quartiles in the context of validation clients remain larger than 66\%.  \figref{fig:Cohesion} also  provides evidence  that the  degradation  of  cohesion  values for each inferred pattern  is  much  more  visible  for \davidtechnique.

In summary, we can say that almost all detected usage patterns achieved by \ourtechnique retain their  informative criteria. Precisely, 75\% of detected usage patterns, according to the boxplot's lower quartile, are characterized with a high usage cohesion in both training and validation contexts.

\textbf{\underline{Study 3.B: Library recommendation.}}
\tabref{tab:recall} reports the recall rate@$k$ for both \ourtechnique and
\davidtechnique while varying the value of $k \in \{1,3,5,7,10\}$. We notice that, as expected, larger $k$ values achieve higher recall rates for both approaches. More specifically, we can see that when comparing the recall rate,  \ourtechnique performs clearly better in terms of recall@1 and recall@3. However, the starting from $k=5$, \davidtechnique tends to achieve  better results. This indicates that \ourtechnique is more efficient in recommending correct libraries in the top ranks within the recommendation list when using \ourtechnique. The MRR results support this observation with a score of 0.15 for \ourtechnique. 


Conversely, good recommendations are achieved for more targeted client systems when using 
\davidtechnique. This is mainly due to the fact that \davidtechnique's patterns are mainly composed of utility libraries, unlike the \ourtechnique's patterns which are mainly composed of domain specific libraries. However, one can notice that recommending utility libraries that are commonly used is less useful in practice.
It is also worth mentioning   that  for each fold, due to the large number of systems (3,800 validation client systems) that require library recommendations, the recall values achieved by both \ourtechnique and \davidtechnique are still low (below 0.34) as reported in \tabref{tab:recall}. 

\begin{table}[]
\tabcolsep=0.5cm
\centering
\caption{Recommendation recall rate and MRR results achieved by both \ourtechnique and \davidtechnique.
\label{tab:recall}}
\begin{tabular}{@{}lcc@{}}
\toprule
          & \textbf{\ourtechnique} & \textbf{LibRec} \\ \midrule
Recall@1  & 0.12                   & 0.01            \\
Recall@3  & 0.14                   & 0.11            \\
Recall@5  & 0.15                   & 0.19            \\
\end{tabular}
\vspace{-0.3cm}
\end{table}


%% file: sections/Discussion.tex
\section{Discussion}\label{sec:discussion}

We applied our approach to over 6,000 popular third-party libraries and 38,000 client systems in order to detect possible library usage patterns. The detected patterns should be informative to help developers in automatically discovering existing library sets  and therefore relieve the developers from the burden of doing so manually. More interestingly, our variant $\epsilon$-DBSCAN shows high scalability and performance with our large dataset.

The evaluation of our approach took into account the potential generalization of the identified patterns to other client systems and showed that these usage patterns remain informative for other clients. 
One of the key contributions of this work is the adaptation of our variant $\epsilon$-DBSCAN algorithm for mining library usage patterns. We have opted for this DBSCAN-based technique rather than a standard clustering technique since DBSCAN has the notion of noise, and it widely considered robust to outliers.

The application of our technique to detect library usage patterns requires the setting of thresholds that may impact its output. 
For instance, the $maxEpsilon$ parameter in the clustering algorithm controls the cohesion (PUC) strength of the detected patterns. 
A small value leads to highly cohesive clusters which means that the detected patterns are more informative. Hence, decreasing the value of this parameter 
would result in an improvement in cohesion of the detected patterns. However, in this case the number and the consistency (generality) of the detected patterns could decrease because the highly 
cohesive detected patterns may not be shared by a large number of clients. To
avoid bothering potential users of our approach with tuning the value of $maxEpsilon$,  we set it to a default value of 0.5 
which ensures that the libraries within patterns are at least used more frequently together than separately.

To get more qualitative sense, we describe one of the inferred patterns
identified by our technique that can fulfil the requirements of the case
scenario discussed in \Secref{sec:sakaiproject} and that provide useful libraries
for potential extensions of the system. The developer would use {\tt scheduler}{\tt
-api} and {\tt mailsender-api} rather than the {\tt quartz} and {\tt
commons-email}. This pattern has different cohesion layers, and provides at the first layer, the libraries {\tt sakai} {\tt
-calendar-api}
 and {\tt sakai-presence-api}
  . In the second layer, we find three libraries that are added to the pattern, namely {\tt portal-chat}
 , {\tt messageforums-tool}
  and {\tt mailsender}{\tt -api}
 . At the external usage cohesion layer the {\tt scheduler-api} is added to the
pattern. Indeed, these libraries have been frequently co-used in a set of 18 client systems at least in our dataset.

It is worth noticing   that we found a trade-off between the usage cohesion of the detected patterns and their generalization. Indeed, another example of more generalizable patterns that was inferred 
when $maxEpsilon$ parameter reached a relatively high value is the one formed in his core layer with some libraries of the Spring framework such 
as {\tt spring-beans }
, {\tt spring-context}
and {\tt spring-orm}
. Then, in the second layer, we find some libraries of the Hibernate framework  such  as the  {\tt hibernate-} {\tt entitymanager}
and {\tt hibernate-annotations}
 Finally, in the third layer, we find some json libraries  such us the  {\tt jackson-} {\tt databind }
 . The pattern continue growing until including some utility libraries of logging and testing. These libraries have been co-used in a set of hundreds of client systems.


%% file: sections/RelatedWork.tex
\section{Related Work}\label{sec:relatedWork} 
Recently, different aspects around library usage have gained considerable
attentions.
Existing contributions can be organized into different categories according to the purpose of their proposed techniques: (\textit{i}) third-party library usage, (\textit{ii}) code completion, (\textit{iii}) library API usage example, (\textit{iv}) API usage visualization, (\textit{v}) exploration of API usage obstacles, and (\textit{vi}) mining API usage patterns.

\paragraph*{\it Third-party Usage at the Library Level}
Several work were interested in the third-party usage at the library level, but
for different purpose such as library refactoring \cite{penta2002knowledge},
library recommendation \cite{david13} and
library miniaturization \cite{antoniol,antoniol2003moving}. The most related to our work is the one by Thung
\etal~\cite{david13}. In this work the authors proposed a hybrid approach that combines association  rule mining and  collaborative  filtering to recommends libraries based on  their usage on similar client systems. This approach is most related to our in the sense that it is also based on library usage through their client systems dependency. However, both approaches have different purpose. Thung \etal~\cite{david13} are interested in similarity between client systems to recommend libraries based on dependency  of the  similar clients, whereas in our case we are interested in the overall libraries usage to discover multi-layers library patterns.

\paragraph*{\it Code Completion} 
Enhancing current completion systems to work more effectively with large APIs
have been investigated in
\cite{Nguy12,Bruch09,hou2011evaluation,McMi11,endrikat2014api}.
This body of work makes use of database of API usage recommendation, type hierarchy, context filtering and API methods functional roles for improving the performance of API method call completion. 
Recently, Asaduzzaman \etal \cite{asad2014} proposed a context sensitive code completion technique that uses, in addition to the aforementioned information, the context of the  method call. 

\paragraph*{\it API Usage Example}  
A similar body of work is interested in example recommendation of API usage \cite{Wang11,Ekok11,Buse12,montandon2013documenting}. 
Existing contributions can be organized in two groups: 
IDE-based recommendation systems and JavaDoc-based recommendation systems. 
These contributions tried to instruments API documentation with usage examples 
based on a static slicing, clustering and pattern abstraction.

\paragraph*{\it API Usage Visualization} 
Other contributions tried to enhance understanding API usage through explorative
and interactive methods
\cite{Moritz13:6693127,kula2014visualizing,parnin2012crowd,de2013multi,saied2015visualization}.
This body of work described multi-dimensional exploration of API usage. 
The explored dimensions are related to the hierarchical organization of projects and APIs, metrics of API usage and API domains.
A visualisation strategy would necessarily enricher the usefulness of our
approach.

\paragraph*{\it Exploration of API Usage Obstacles} 
From another perspective, the work in
\cite{hou2011obstacles,wang2013detecting,SaiedSanerConstraintDoC2015} explored API usage obstacles through analyzing developers questions in  Q\&A website. 
This allows API designers to understand the problems faced while using their API, and to make corresponding improvements.

\paragraph*{\it Mining API Usage Patterns}  
Other contributions related  to ours are those interested
in mining API usage patterns
\cite{Wang:2013:MSH:2487085.2487146,Uddi12,Zhon09,li2005pr,saied2016cooperative}.
These contributions adopted different categories of API usage patterns, different techniques for inferring patterns and different ways to assess patterns correctness and usefulness. 
The most prominent categories are temporal (\cite{Uddi12}), unordered
(\cite{li2005pr}) and sequential (\cite{Wang:2013:MSH:2487085.2487146,Zhon09})
usage patterns. These categories were assessed through consistency, coverage and
succinctness of the mined usage patterns. Zhong \etal~\cite{Zhon09} developed the MAPO tool for mining sequential API
  usage patterns. MAPO clusters frequent API
  method call sequences extracted from code snippets, based on the number of
  called API methods and textual similarity of class and method names between
  different snippets. Our  work consider a different  level  of  granularity 
  from  these   existing works. Previous approaches infer API usage patterns at the API element level (i.e. methods call). 
  In our case we are interested in inferring usage pattern at the granularity of entire library. Past approaches assume that the developer already selected the relevant 
  library and he only need to learn how to use the methods in this library. Our work does not make  this  assumption,  and  thus  complements  the existing studies. Indeed, our approach could be used as an early phase to infer set of library consistently co-used together. Then existing approaches could be applied to learn how to use particular methods within the patterns' libraries.

%% file: sections/Conclusion.tex
Third-party library reuse has  become vital in modern software development. The number of libraries provided on the Internet is exponentially growing which would provide several reuse opportunities.
 In this paper, we introduced an automated approach to detect multi-level library usage patterns -- a collection of libraries that are commonly used together by client systems, distributed through multiple levels of cohesion. To this end, we adopted a variant of the standard clustering algorithm DBSCAN, namely $\epsilon$-DBSCAN especially for the library usage patterns detection. We evaluated our approach on large dataset of 6,638 popular libraries from Maven repository, and a large population of 38,000 client systems from Github, and we compared its results to those of a state-of-the-art approach. 
The results indicate that our approach gives a comprehensive overview on third-party library usage patterns. The obtained usage patterns exhibit high usage cohesion with an average of 77\% , and could be generalizable to other systems. Automatically detecting library usage patterns would support developers in enhancing the library space discovery, and attract their attention the missed reuse opportunities.

As future work, we are planning  to provide an Eclipse plug-in to automatically notify developers with potential library usage patterns that are related to their currently used libraries and implementation. 
Furthermore, we are planing to unify our library-level usage pattern detection with method-level usage pattern detection techniques in order to provide a comprehensive package for developers supporting them in understanding and reusing third-party libraries.